\documentclass[twocolumn,pra,aps,amsmath]{revtex4}

\usepackage{graphics}
\usepackage{epsfig}

\begin{document}
\newcommand*{\ket}[1]{$|{#1}\rangle$}
\newcommand*{\beq}{\begin{equation}}
\newcommand*{\eeq}{\end{equation}}
\newcommand*{\beqs}{\begin{equation*}}
\newcommand*{\eeqs}{\end{equation*}}

\title{Electron interferometry with nano-gratings}
\author{Alexander D. Cronin}
\author{Ben McMorran}
 \affiliation{Department of Physics, University of Arizona, Tucson, AZ 85721}

\begin{abstract}
We present an electron interferometer based on near-field
diffraction from two nanostructure gratings. Lau fringes are
observed with an imaging detector, and revivals in the fringe
visibility occur as the separation between gratings is increased
from 0 to 3 mm. This verifies that electron beams diffracted by
nanostructures remain coherent after propagating farther than the
Talbot length $z_T = 2d^2/\lambda = 1.2$ mm, and hence is a proof
of principle for the function of a Talbot-Lau interferometer for
electrons. Distorted fringes due to a phase object demonstrates an
application for this new type of electron interferometer.
\end{abstract}

\date{\today}
\maketitle

\begin{figure}
\includegraphics[width = 8cm]{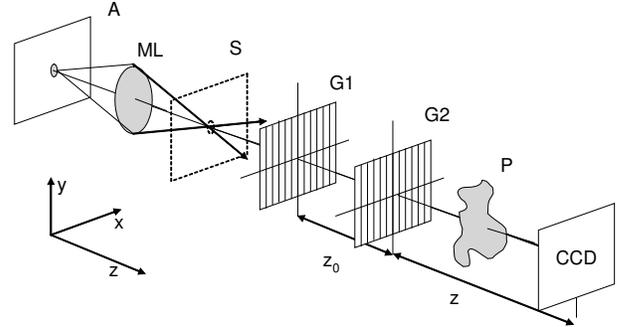}
\caption{Lau interferometer for electrons.  A, 100-$\mu$m
aperture; ML, magnetic lens; S, the electron beam converges to a
10 $\mu$m spot in this plane; G1, nano-grating; G2, 2nd
nano-grating; P, phase object; CCD, imaging screen. In our
experiments $z_0$ is in the range 0-3 mm, and $z$=80 cm. The
distance between S and G1 is typically 2 cm, and the divergence of
the beam from S is 5 $\times 10^{-3}$ radians. Not shown:  the
thermionic tungsten filament, grid cap, and condenser lens are
located before A.}\label{fig:setup}
\end{figure}

Near-field interference effects that result in self-similar images
of a periodic structure were noticed by Talbot in 1836, and later
described as Fourier images \cite{TAL36,COM57a,PAT89}. One
remarkable feature is that revivals in image visibility occur
periodically as the plane of observation is separated from the
periodic structure. The visibility of self-images is maximized at
half-integer multiples of the Talbot distance $z_T =
2d^2/\lambda$, with $d$ being the period of the structure and
$\lambda$ the wavelength of the light or de Broglie waves
illuminating the structure. Partially coherent waves are required
to observe self-images of a single grating (the Talbot effect).
However, a related phenomenon (the Lau effect) occurs with
incoherent light if \emph{two} gratings are used \cite{PAT89,
LAU48,JAL79}. Fringes are then formed behind the second grating,
and the fringe visibility oscillates as a function of grating
separation. These Lau fringes have maximum visibility on a distant
screen when the gratings are separated by $n d^2 / \lambda$, with
$n$ being an integer. In a Talbot-Lau interferometer, Lau fringes
are detected with the aid of a third grating, but the fringes can
also be observed directly on a screen, thus making a Lau
interferometer as shown in Figure \ref{fig:setup}.

\begin{figure}
\includegraphics[width = 5cm]{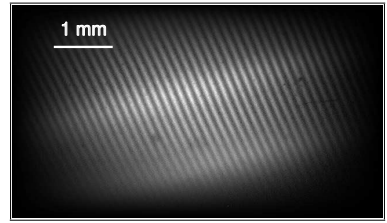}
\caption{Lau fringes formed with 5 keV electrons and two 100-nm
period gratings separated by 0.6 mm (half the Talbot length). This
figure was obtained by imaging a phosphor screen with a CCD.
}\label{fig:fringes}
\end{figure}

\begin{figure}
\includegraphics[width = 8.2cm]{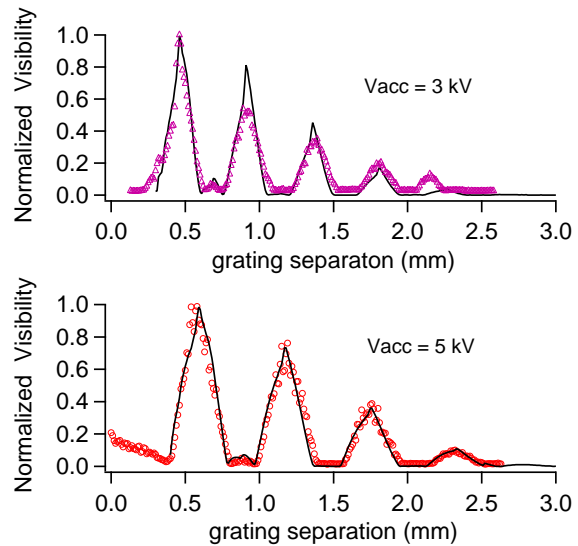}
\caption{Revivals in fringe visibility as a function of grating
separation ($z_0$). Data (symbols) are compared to theory (line)
from reference \cite{BAZ03} for two gratings each with 35\% open
fraction.  Visibility maxima are found when the gratings are
separated by a half integer number of Talbot lengths, and this
depends on the electron accelerating voltage (Vacc). The maximum
visibility is 30\% for both the 3keV and 5keV electrons.
Visibility decays when the fringe period is comparable to the
resolution of the detector.}\label{fig:revivals}
\end{figure}

Here we present a Lau interferometer for electrons based on two
nanostructure gratings that each have a period of $d=100$ nm. With
medium energy (5 keV) electrons that have a de Broglie wavelength
of $\lambda$ = 17 pm, the Talbot length is 1.16 mm. An imaging
detector 80 cm beyond the gratings was used to observe the Lau
fringes shown in Figure \ref{fig:fringes}, and the fringe
visibility as a function of grating separation is plotted in
Figure \ref{fig:revivals}.  If the fringes are analyzed with a
third grating (even a digital grating in the image processing can
achieve this purpose), then this apparatus serves as a Talbot-Lau
interferometer. However, even more information is gained by
studying images of the Lau fringes directly.

Interferometers based on the Talbot and Lau effects have found
applications in light optics \cite{PAT89,JAL79,SIL72}, in atom
optics \cite{CEH95,NKP97,CLL94,BHU02,BAZ03}, and more recently
with x-rays \cite{DNS02}.  Yet even though electron interferometry
is a mature field \cite{TON87,TON94,TON99}, neither Lau nor
Talbot-Lau interferometer designs have been operated with
electrons until now.

Perhaps the chief reason that Talbot-Lau interferometers have not
previously been created for electrons is that suitable periodic
structures have not been available. Crystals with a lattice period
on the order of 1 nm can serve as a grating, but the resulting
Talbot length of 200 nm is too short for many practical
interferometer experiments \cite{note2}. A further complication is
that the angular misalignment of the two gratings must be smaller
than one grating period over the height of the beam. Hence
alignment within $10^{-6}$ radians is required when using crystal
gratings and a 1 mm high beam. These limitations are overcome by
using nanostructure gratings with a 100 nm period. Then the talbot
length is increased to 1 mm, and the alignment tolerance is
relaxed to $10^{-3}$ radians.

Nanostructures for electron optics are new. The gratings that we
use are fabricated at MIT \cite{SSS95}, and were only recently
used to study electron diffraction \cite{GBB05,MPS06}. Similar
gratings have enabled several atom interferometers
\cite{KET91,CAM91,CLL94,BHU02}, and thicker gratings have given
first light to x-ray Talbot interferometers \cite{DNS02}. That
nano-structures can be used for coherent electron optics was not
obvious for several reasons. Local charging of the
nano-structures, non-uniform image charge interactions with the
nano-structures, and the possibility of coulomb drag and electron
energy loss due to passing within a few nanometers of a surface
are just a few mechanisms that could lead to decoherence.  Our
results show that sufficient coherence is maintained to operate an
electron interferometer with nanostructure gratings.

In the rest of this paper we describe the electron optics setup,
and the diffraction theory used to study the revivals in Figure
\ref{fig:revivals}.  We comment on the role of image charge
interactions between electrons and the nano-gratings. Then we
demonstrate an application of this interferometer: the study of
the index of refraction for electrons due fields around a charged
needle tip.

The Lau fringes have highest visibility when the grating
separation ($z_0$) and the distance to the screen ($z$) satisfy
\begin{equation}
\frac{z_0 z}{z_0 + z} = \frac{n d^2 }{\lambda}.  \label{eq:z_0}
\end{equation}
Then the period of the Lau fringes is
\begin{equation}
d' = d\frac{z_0 + z}{z_0}.   \label{eq:d'}
\end{equation}
where $d$ is the period of the gratings. Equations \ref{eq:z_0}
and \ref{eq:d'} are derived in the Fresnel approximation in
references \cite{SIL72,PAT89,BAZ03}. Because the distance to the
screen is typically 800 times as long as the separation between
gratings ($z/z_0 = 800$) in our experiment, the fringes are
effectively magnified so that $d' = 800 d$.  The ratio of periods
for G1, G2 and the detected fringes is therefore 1:1:800.

\begin{figure}
\includegraphics[width = 9cm]{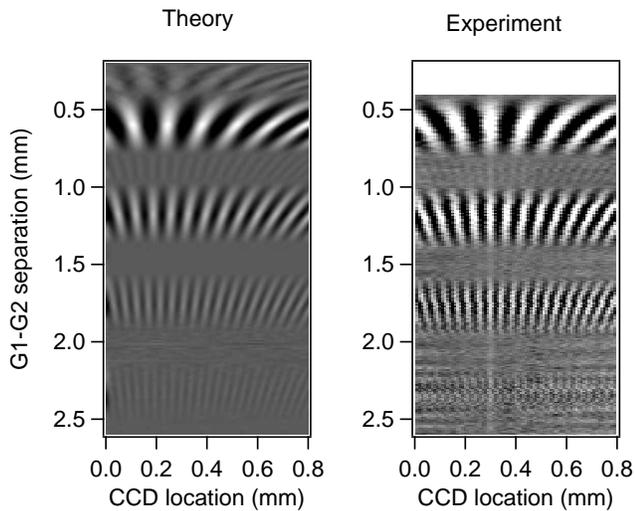}
\caption{Lau fringes vs. grating separation.  Each row in this
composite image represents a row from a different raw image. The
raw images were taken with different values of G1-G2 separation
($z_0$).  The theoretical composite was generated with the theory
given in reference \cite{BAZ03}. A point spread function for the
imaging detector (Gaussian with rms = 40 $\mu$m), and the open
fraction of the gratings (35\%) is included in the theory.}
\label{fig:composite}
\end{figure}

To describe in detail the shape and visibility, \mbox{$V  \equiv
(max-min)/(max + min)$}, of the fringes as a function of grating
separation we relied on a calculation given in Equation 3 of
reference \cite{BAZ03}. Briefly, the concept is that the first
grating is considered to be a set of mutually incoherent point
sources of monochromatic waves.  Waves from each point source are
diffracted by G2 and arrive at the imaging screen in the
near-field (Fresnel) diffraction regime. The intensity pattern on
the screen is generated by adding diffraction patterns from each
point source.  Grating G1 and the screen define the position of
many sources and detectors; they function as classical components.
In contrast, grating G2 acts as a quantum mechanical diffraction
object.  This theory was used to generate the curves in Figure
\ref{fig:revivals} and the theoretical portion of Figure
\ref{fig:composite}.

We are justified in considering incoherent illumination of G1 by
the van Cittert-Zernike theorem, which gives the transverse
coherence length, $\ell_{tcoh} \approx \lambda / \theta$ in terms
of $\theta$, the angle subtended by the source. The electron beam
is focused to a 10 $\mu$m spot just 2 cm in front of G1 (see
figure \ref{fig:setup}) and the spot size in this plane cannot be
made smaller because of the extended electron source and lens
abberations. This corresponds to a coherence length of
$\ell_{tcoh} = 10$ nm, which is smaller than the grating period of
$d=100$ nm.

As remarked in reference \cite{BAZ03}, the essential coherent
effect is diffraction at the second grating.  The first grating
simply serves as an array of sources, and the fringes could be
detected by a transmission mask (with a period $d'$) and an
integrating detector.   One virtue of using an imaging detector,
with a digital sampling density of 100 pixels per mm, is that we
can study fringes of varying period ($d'$) as the grating
separation ($z_0$) is changed.  Images also reveal the
\emph{fractional Talbot effect} by showing fringes of half the
period compared to that given by Equation \ref{eq:d'}, when the
grating separation is three quarters of the Talbot length, $z_0 =
3/4 z_T$.  The fractional Talbot effect was also observed with
$z_0 = 1/4 z_T$.

Our experimental data are best fit by a grating transmission
function for G1 and G2 that is described by a 35\% open fraction,
and a weak image-charge interaction between electrons and the
grating bars.   Image-charge interactions were discussed in detail
in \cite{GBB05,MPS06}, and the have a similar effect on the
electron optics as the Casimir-Polder interaction does for atom
optics \cite{BAZ03,GST99,PCS05}.  We include the strength of the
image-charge and the physical size of the grating windows as two
free parameters in the transmission function for the second
grating. The best fit image charge for 3 keV electrons is $q' =
0.03 |e|$, and the best fit image charge for 5 keV electrons is
only $q' = 0.01 |e|$. This suggests that the electron-surface
interaction at the nanometer scale depends on the velocity of
electrons with respect to the surface.  More research to
investigate this is under way.

\begin{figure}
\includegraphics[width = 8cm]{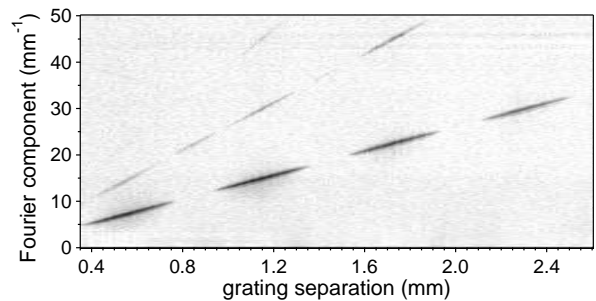}
\caption{Power spectra of Lau fringes vs. grating separation. Each
column is given by the Fourier transform of a different row of
data in Figure \ref{fig:composite}.}
 \label{fig:fft}
\end{figure}

Additional analysis of the shape of the fringes was accomplished
by taking the Fourier transform of images such as Figure
\ref{fig:fringes}. These image transforms show how the spatial
frequency of the fringes change with grating separation.  A
composite image of fringe spectra, in which each column represents
the spatial frequency spectrum of fringes for a specific grating
separation $(z_0)$, is shown in Figure \ref{fig:fft}.  The
fractional Talbot effect is clear at the grating separation $z_0 =
0.85$ mm. The higher harmonics of spatial frequency indicate that
the fringes are not purely sinusoidal, but tend in places to look
more like the binary (Ronchi rule) masks made by ideal absorbing
nanostructure gratings. This is the self-imaging property of the
Talbot and Lau effects.

\begin{figure}
\includegraphics[width = 8cm]{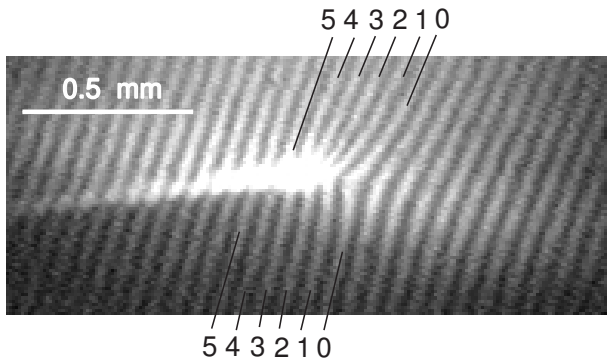}
\caption{Distorted fringes around the shadow of a charged wire.
Consecutive fringes are labelled on the top and bottom to
emphasize the discontinuity that arises due to the wire acting as
a bi-prism. The 100 $\mu$m diameter wire was held at 4.5 volts,
and surrounded by a grounded cylinder of radius 3 cm.}
 \label{fig:wire}
\end{figure}

To demonstrate that this interferometer can be used to study phase
shifts for electrons due to various objects, we inserted a charged
wire in the beam after the second grating in plane $\texttt{P}$ of
Figure \ref{fig:setup}. Like a lightening rod, the tip of the wire
causes large gradients in electric potential. The electric
potential in the space around the wire changes the index of
refraction for electron waves and distorts the interference
fringes as shown in Figure \ref{fig:wire}.

Multiple paths through the interferometer sample different parts
of the phase object, therefore we are sensitive only to gradients
in phase, not phase directly.  This design is known as a shearing
interferometer and shifts in fringe position are proportional to
$[\partial/\partial x] \Phi(x,y)$ and distortion in the fringes is
associated with $[\partial/\partial x]^2 \Phi(x,y)$, where
$\Phi(x,y) = \int n(x,y,z) k_0 dz$, with $n(x,y,z)$ being the
index of refraction
\cite{PAT89,SIL72,JAL79,TON87,TON99,TON94,DNS02}. For electron de
Broglie waves this is $\Phi(x,y)=
 \int \sqrt{ (2m/\hbar^2) [ E + |e|V(x,y,z) ] } dz $ where $V(x,y,z)$
is the electric potential, $E$ is the incident energy of the
electrons, $m$ the electron mass, $e$ the magnitude of the
electron charge, and $\hbar$ is Planck's constant divided by
$2\pi$.

Because phase gradients are caused by energy gradients, i.e.
forces, the sensitivity demonstrated here is not different in
principle from a deflection sensor that can study classical forces
on electrons.  However, the virtue of the interferometer is still
apparent because the fringes make it easier to detect deflections.
One advantage is that the fringe period is 10 times finer than the
smallest focus that could be produced with the incoherent beam and
abberations of the lens system in this apparatus. Second, the
fringes are formed simultaneously over a large spatial region
without scanning the beam.

A continuous wire produces a uniform linear phase gradient
($\nabla \Phi(x,y)$) with opposite sign on either side of the
wire.  That is how it serves as a bi-prism \cite{TON99}.  Around
the tip of the freely suspended charged wire, however, there is a
strong double gradient term ($\nabla^2 \Phi(x,y)$).  Hence fringe
distortion is expected around the tip of the charged wire, and
uniform fringe shifts are expected along the sides of the charged
wire.  Since the shifts are perpendicular to the wire, and in
general the wire can be skew to the grating bars, fringes on
either side of the wire can appear out of step as indicated in
figure \ref{fig:wire}.  This serves as a proof of principle that
the interferometer setup presented here can be used to study
differential phase shifts due to a phase object.

At the time of writing of this manuscript, we are aware of the
construction of a Mach-Zehnder electron interferometer in the
group of H. Batelaan, which requires a much higher degree of
electron spatial coherence than our interferometer.  Our results
are distinct in that we use an incoherent electron beam and we
image electron interference fringes directly. The imaging tool
allows us to detect fringes with arbitrary period ($d'$), and thus
permitted us to study quantitatively the revivals in fringe
visibility as a function of grating separation.  Imaging also
enabled us to observe the fractional Talbot effect with electrons
and nanostructures.  The most useful result of using an imaging
detector is the ability to study fringe distortions due to phase
objects in the electron interferometer.

In conclusion, we demonstrated a novel electron interferometer
that uses two nanostructure gratings and near-field interference
effects.  We demonstrated the Lau effect for electrons and
observed revivals in fringe visibility when the gratings are
separated by multiples of the Talbot length. This type of electron
interferometer does not require spatially coherent electron waves
from the electron gun, but still it tests how well the
nanostructures generate and preserve coherence for electron waves.
The effect of image-charge interactions between electrons and the
grating structure was observed, but it does not inhibit the
electron interference. The apparatus is a rudimentary sheering
interferometer, and serves to demonstrate differential phase
shifts. We have thus shown that metal-coated silicon nitride
nanostructure gratings can be used as elements for coherent
electron optics.

The authors acknowledge D. Bentley for the electron gun, and Mark
Robertson-Tessi for technical assistance. This work was supported
by the National Science Foundation Grant No. 0354947, and No.
0526954.



\begin{thebibliography}{23}
\expandafter\ifx\csname
natexlab\endcsname\relax\def\natexlab#1{#1}\fi
\expandafter\ifx\csname bibnamefont\endcsname\relax
  \def\bibnamefont#1{#1}\fi
\expandafter\ifx\csname bibfnamefont\endcsname\relax
  \def\bibfnamefont#1{#1}\fi
\expandafter\ifx\csname citenamefont\endcsname\relax
  \def\citenamefont#1{#1}\fi
\expandafter\ifx\csname url\endcsname\relax
  \def\url#1{\texttt{#1}}\fi
\expandafter\ifx\csname
urlprefix\endcsname\relax\def\urlprefix{URL }\fi
\providecommand{\bibinfo}[2]{#2}
\providecommand{\eprint}[2][]{\url{#2}}

\bibitem[{\citenamefont{Talbot}(1836)}]{TAL36}
\bibinfo{author}{\bibfnamefont{H.~F.} \bibnamefont{Talbot}},
  \bibinfo{journal}{Philos. Mag.} \textbf{\bibinfo{volume}{9}},
  \bibinfo{pages}{401} (\bibinfo{year}{1836}).

\bibitem[{\citenamefont{Cowley and Moodie}(1957{\natexlab{a}})}]{COM57a}
\bibinfo{author}{\bibfnamefont{J.~M.} \bibnamefont{Cowley}} \bibnamefont{and}
  \bibinfo{author}{\bibfnamefont{A.~F.} \bibnamefont{Moodie}},
  \bibinfo{journal}{Proc. Phys. Soc. London}
  \textbf{\bibinfo{volume}{70}}, \bibinfo{pages}{486}
  (\bibinfo{year}{1957}{\natexlab{a}});
  \textbf{\bibinfo{volume}{70}}, \bibinfo{pages}{497}
  (\bibinfo{year}{1957}{\natexlab{b}});
  \textbf{\bibinfo{volume}{70}}, \bibinfo{pages}{505}
  (\bibinfo{year}{1957}{\natexlab{c}});
  \textbf{\bibinfo{volume}{76}}, \bibinfo{pages}{378} (\bibinfo{year}{1960}).

\bibitem[{\citenamefont{Patorski}(1989)}]{PAT89}
\bibinfo{author}{\bibfnamefont{K.}~\bibnamefont{Patorski}},
  \bibinfo{journal}{Progress in Optics} \textbf{\bibinfo{volume}{27}},
  \bibinfo{pages}{3} (\bibinfo{year}{1989}).

\bibitem[{\citenamefont{Lau}(1948)}]{LAU48}
\bibinfo{author}{\bibfnamefont{E.}~\bibnamefont{Lau}}, \bibinfo{journal}{Ann.
  Phys.} \textbf{\bibinfo{volume}{6}}, \bibinfo{pages}{417}
  (\bibinfo{year}{1948}).

\bibitem[{\citenamefont{Jahns and Lohmann}(1979)}]{JAL79}
\bibinfo{author}{\bibfnamefont{J.}~\bibnamefont{Jahns}} \bibnamefont{and}
  \bibinfo{author}{\bibfnamefont{A.~W.} \bibnamefont{Lohmann}},
  \bibinfo{journal}{Opt. Comm.} \textbf{\bibinfo{volume}{28}},
  \bibinfo{pages}{263} (\bibinfo{year}{1979});
\bibinfo{author}{\bibfnamefont{H.~O.} \bibnamefont{Bartelt}} \bibnamefont{and}
  \bibinfo{author}{\bibfnamefont{J.}~\bibnamefont{Jahns}},
  \bibinfo{journal}{Opt. Comm.} \textbf{\bibinfo{volume}{30}},
  \bibinfo{pages}{268} (\bibinfo{year}{1979}).

\bibitem[{\citenamefont{Silva}(1972)}]{SIL72}
\bibinfo{author}{\bibfnamefont{D.~E.} \bibnamefont{Silva}},
  \bibinfo{journal}{Applied Optics} \textbf{\bibinfo{volume}{11}},
  \bibinfo{pages}{2613} (\bibinfo{year}{1972}).

\bibitem[{\citenamefont{Chapman et~al.}(1995)\citenamefont{Chapman, Ekstrom,
  Hammond, Schmiedmayer, Tannian, Wehinger, and Pritchard}}]{CEH95}
\bibinfo{author}{\bibfnamefont{M.~S.} \bibnamefont{Chapman}},
  \bibinfo{author}{\bibfnamefont{C.~R.} \bibnamefont{Ekstrom}},
  \bibinfo{author}{\bibfnamefont{T.~D.} \bibnamefont{Hammond}},
  \bibinfo{author}{\bibfnamefont{J.}~\bibnamefont{Schmiedmayer}},
  \bibinfo{author}{\bibfnamefont{B.~E.} \bibnamefont{Tannian}},
  \bibinfo{author}{\bibfnamefont{S.}~\bibnamefont{Wehinger}}, \bibnamefont{and}
  \bibinfo{author}{\bibfnamefont{D.~E.} \bibnamefont{Pritchard}},
  \bibinfo{journal}{Phys. Rev. A} \textbf{\bibinfo{volume}{51}},
  \bibinfo{pages}{R14} (\bibinfo{year}{1995}).

\bibitem[{\citenamefont{Nowak et~al.}(1997)\citenamefont{Nowak, Kurtsiefer,
  Pfau, and David}}]{NKP97}
\bibinfo{author}{\bibfnamefont{S.}~\bibnamefont{Nowak}},
  \bibinfo{author}{\bibfnamefont{C.}~\bibnamefont{Kurtsiefer}},
  \bibinfo{author}{\bibfnamefont{T.}~\bibnamefont{Pfau}}, \bibnamefont{and}
  \bibinfo{author}{\bibfnamefont{C.}~\bibnamefont{David}},
  \bibinfo{journal}{Opt. Lett.} \textbf{\bibinfo{volume}{22}},
  \bibinfo{pages}{1430} (\bibinfo{year}{1997}).

\bibitem[{\citenamefont{Clauser and Li}(1994)}]{CLL94}
\bibinfo{author}{\bibfnamefont{J. F.}~\bibnamefont{Clauser}} \bibnamefont{and}
  \bibinfo{author}{\bibfnamefont{S.}~\bibnamefont{Li}}, \bibinfo{journal}{Phys.
  Rev. A} \textbf{\bibinfo{volume}{49}}, \bibinfo{pages}{R2213}
  (\bibinfo{year}{1994}).

\bibitem[{\citenamefont{Brezger et~al.}(2002)\citenamefont{Brezger,
  Hackermuller, Uttenthaler, Petschinka, Arndt, and Zeilinger}}]{BHU02}
\bibinfo{author}{\bibfnamefont{B.}~\bibnamefont{Brezger}},
  \bibinfo{author}{\bibfnamefont{L.}~\bibnamefont{Hackermuller}},
  \bibinfo{author}{\bibfnamefont{S.}~\bibnamefont{Uttenthaler}},
  \bibinfo{author}{\bibfnamefont{J.}~\bibnamefont{Petschinka}},
  \bibinfo{author}{\bibfnamefont{M.}~\bibnamefont{Arndt}}, \bibnamefont{and}
  \bibinfo{author}{\bibfnamefont{A.}~\bibnamefont{Zeilinger}},
  \bibinfo{journal}{Phys. Rev. Lett.} \textbf{\bibinfo{volume}{88}},
  \bibinfo{pages}{479} (\bibinfo{year}{2002}).

\bibitem[{\citenamefont{Brezger et~al.}(2003)\citenamefont{Brezger, Arndt, and
  Zeilinger}}]{BAZ03}
\bibinfo{author}{\bibfnamefont{B.}~\bibnamefont{Brezger}},
  \bibinfo{author}{\bibfnamefont{M.}~\bibnamefont{Arndt}}, \bibnamefont{and}
  \bibinfo{author}{\bibfnamefont{A.}~\bibnamefont{Zeilinger}},
  \bibinfo{journal}{Journal of Optics B}
  \textbf{\bibinfo{volume}{5}}, \bibinfo{pages}{S82} (\bibinfo{year}{2003}).

\bibitem[{\citenamefont{David et~al.}(2002)\citenamefont{David, Nohammer,
  Solak, and Ziegler}}]{DNS02}
\bibinfo{author}{\bibfnamefont{C.}~\bibnamefont{David}},
\bibinfo{author}{\textit{et al}.},
  \bibinfo{journal}{App. Phys. Lett.} \textbf{\bibinfo{volume}{81}},
  \bibinfo{pages}{3287} (\bibinfo{year}{2002});
\bibinfo{author}{\mbox{\bibfnamefont{A.}~\bibnamefont{Momose}}},
\bibinfo{author}{\textit{et al}.},
  \bibinfo{journal}{Jap. J. App. Phys.}
  \textbf{\bibinfo{volume}{42}}, \bibinfo{pages}{L866}
  (\bibinfo{year}{2003});
\bibinfo{author}{\bibfnamefont{A.}~\bibnamefont{Momose}},
  \bibinfo{journal}{Jap. J. App. Phys.}
  \textbf{\bibinfo{volume}{44}}, \bibinfo{pages}{6355}
  (\bibinfo{year}{2005});
\bibinfo{author}{\bibfnamefont{T.}~\bibnamefont{Weitkamp}},
\bibinfo{author}{\textit{et al}.},
  \bibinfo{journal}{App. Phys. Lett.} \textbf{\bibinfo{volume}{86}}
  (\bibinfo{year}{2005}).

\bibitem[{\citenamefont{Tonomura}(1987)}]{TON87}
\bibinfo{author}{\bibfnamefont{A.}~\bibnamefont{Tonomura}},
  \bibinfo{journal}{Rev. Mod. Phys.} \textbf{\bibinfo{volume}{59}},
  \bibinfo{pages}{639} (\bibinfo{year}{1987}).

\bibitem[{\citenamefont{Tonomura et~al.}(1995)\citenamefont{Tonomura, Allard,
  Pozzi, Joy, and Ono}}]{TON94}
\bibinfo{editor}{\bibfnamefont{A.}~\bibnamefont{Tonomura}},
  \bibinfo{editor}{\bibfnamefont{L.~F.} \bibnamefont{Allard}},
  \bibinfo{editor}{\bibfnamefont{G.}~\bibnamefont{Pozzi}},
  \bibinfo{editor}{\bibfnamefont{D.~C.} \bibnamefont{Joy}}, \bibnamefont{and}
  \bibinfo{editor}{\bibfnamefont{Y.~A.} \bibnamefont{Ono}}, eds.,
  \emph{\bibinfo{title}{Proceedings of the International Workshop on Electron
  Holography}} (\bibinfo{publisher}{Elsevier}, \bibinfo{year}{1995}).

\bibitem[{\citenamefont{Tonomura}(1999)}]{TON99}
\bibinfo{author}{\bibfnamefont{A.}~\bibnamefont{Tonomura}},
  \emph{\bibinfo{title}{Electron Holography}} (\bibinfo{publisher}{Springer},
  \bibinfo{year}{1999}).

\bibitem[{not()}]{note2}
\bibinfo{note}{For this calculation we assumed high energy (100 keV) electrons
  with a de Broglie wavelength of $\lambda$ = 3.9 pm. Lower energy electrons
  make the Talbot length even shorter.}

\bibitem[{\citenamefont{Savas et~al.}(1995)\citenamefont{Savas, Shah,
  Schattenburg, Carter, and Smith}}]{SSS95}
\bibinfo{author}{\bibfnamefont{T.~A.} \bibnamefont{Savas}},
  \bibinfo{author}{\bibfnamefont{S.~N.} \bibnamefont{Shah}},
  \bibinfo{author}{\bibfnamefont{M.~L.} \bibnamefont{Schattenburg}},
  \bibinfo{author}{\bibfnamefont{J.~M.} \bibnamefont{Carter}},
  \bibnamefont{and} \bibinfo{author}{\bibfnamefont{H.~I.} \bibnamefont{Smith}},
\bibinfo{journal}{J. Vac. Sci. Technol. B}
  \textbf{\bibinfo{volume}{13}}, \bibinfo{pages}{2732}
  (\bibinfo{year}{1995});
\bibinfo{author}{\bibfnamefont{T.~A.} \bibnamefont{Savas}},
  \bibinfo{author}{\bibfnamefont{M.~L.} \bibnamefont{Schattenburg}},
  \bibinfo{author}{\bibfnamefont{J.~M.} \bibnamefont{Carter}},
  \bibnamefont{and} \bibinfo{author}{\bibfnamefont{H.~I.} \bibnamefont{Smith}},
\bibinfo{journal}{J. Vac. Sci. Technol. B}
  \textbf{\bibinfo{volume}{14}}, \bibinfo{pages}{4167} (\bibinfo{year}{1996}).

\bibitem[{\citenamefont{Gronniger et~al.}(2005)\citenamefont{Gronniger,
  Barwick, Batelaan, Savas, Pritchard, and Cronin}}]{GBB05}
\bibinfo{author}{\bibfnamefont{G.}~\bibnamefont{Gronniger}},
  \bibinfo{author}{\bibfnamefont{B.}~\bibnamefont{Barwick}},
  \bibinfo{author}{\bibfnamefont{H.}~\bibnamefont{Batelaan}},
  \bibinfo{author}{\bibfnamefont{T.}~\bibnamefont{Savas}},
  \bibinfo{author}{\bibfnamefont{D.}~\bibnamefont{Pritchard}},
  \bibnamefont{and} \bibinfo{author}{\bibfnamefont{A.}~\bibnamefont{Cronin}},
  \bibinfo{journal}{App. Phys. Lett.} \textbf{\bibinfo{volume}{87}}
  (\bibinfo{year}{2005}).

\bibitem[{\citenamefont{McMorran et~al.}(2006)\citenamefont{McMorran,
  Perreault, Savas, and Cronin}}]{MPS06}
\bibinfo{author}{\bibfnamefont{B.}~\bibnamefont{McMorran}},
  \bibinfo{author}{\bibfnamefont{J.~D.} \bibnamefont{Perreault}},
  \bibinfo{author}{\bibfnamefont{T.~A.} \bibnamefont{Savas}}, \bibnamefont{and}
  \bibinfo{author}{\bibfnamefont{A.}~\bibnamefont{Cronin}},
  \bibinfo{journal}{Ultramicroscopy} \textbf{\bibinfo{volume}{106}},
  \bibinfo{pages}{356} (\bibinfo{year}{2006}).

\bibitem[{\citenamefont{Keith et~al.}(1991)\citenamefont{Keith, Ekstrom,
  Turchette, and Pritchard}}]{KET91}
\bibinfo{author}{\bibfnamefont{D.~W.} \bibnamefont{Keith}},
  \bibinfo{author}{\bibfnamefont{C.~R.} \bibnamefont{Ekstrom}},
  \bibinfo{author}{\bibfnamefont{Q.~A.} \bibnamefont{Turchette}},
  \bibnamefont{and} \bibinfo{author}{\bibfnamefont{D.~E.}
  \bibnamefont{Pritchard}}, \bibinfo{journal}{Phys. Rev. Lett.}
  \textbf{\bibinfo{volume}{66}}, \bibinfo{pages}{2693} (\bibinfo{year}{1991}).

\bibitem[{\citenamefont{Carnal and Mlynek}(1991)}]{CAM91}
\bibinfo{author}{\bibfnamefont{O.}~\bibnamefont{Carnal}} \bibnamefont{and}
  \bibinfo{author}{\bibfnamefont{J.}~\bibnamefont{Mlynek}},
  \bibinfo{journal}{Phys. Rev. Lett.} \textbf{\bibinfo{volume}{66}},
  \bibinfo{pages}{2689} (\bibinfo{year}{1991}).

\bibitem[{\citenamefont{Grisenti et~al.}(1999)\citenamefont{Grisenti,
  Schollkopf, Toennies, Hegerfeldt, and Kohler}}]{GST99}
\bibinfo{author}{\bibfnamefont{R.~E.} \bibnamefont{Grisenti}},
  \bibinfo{author}{\bibfnamefont{W.}~\bibnamefont{Schollkopf}},
  \bibinfo{author}{\bibfnamefont{J.~P.} \bibnamefont{Toennies}},
  \bibinfo{author}{\bibfnamefont{C.~C.} \bibnamefont{Hegerfeldt}},
  \bibnamefont{and} \bibinfo{author}{\bibfnamefont{T.}~\bibnamefont{Kohler}},
  \bibinfo{journal}{Phys. Rev Lett.} \textbf{\bibinfo{volume}{83}},
  \bibinfo{pages}{1755} (\bibinfo{year}{1999}).

\bibitem[{\citenamefont{Perreault et~al.}(2005)\citenamefont{Perreault, Cronin,
  and Savas}}]{PCS05}
\bibinfo{author}{\bibfnamefont{J.~D.} \bibnamefont{Perreault}},
  \bibinfo{author}{\bibfnamefont{A.~D.} \bibnamefont{Cronin}},
  \bibnamefont{and} \bibinfo{author}{\bibfnamefont{T.~A.} \bibnamefont{Savas}},
  \bibinfo{journal}{Phys. Rev. A} \textbf{\bibinfo{volume}{71}},
  \bibinfo{pages}{053612} (\bibinfo{year}{2005}).

\end{thebibliography}
\end{document}